\documentclass[aps,prd,floatfix,showpacs]{revtex4}
\usepackage[dvips]{graphicx}
\newcommand{\be}{\begin{equation}}    % * would eliminate numbering
\newcommand{\ee}{\end{equation}}
\newcommand{\bea}{\begin{eqnarray}}
\newcommand{\eea}{\end{eqnarray}}
\newcommand{\bitem}{\begin{itemize}}
\newcommand{\eitem}{\end{itemize}}

\newcommand{\benum}{\begin{enumerate}}
\newcommand{\eenum}{\end{enumerate}}
\newcommand{\bc}{\begin{center}}
\newcommand{\ec}{\end{center}}
\begin{document}
\title{CLASSICAL SCALING SYMMETRY IMPLIES USEFUL NONCONSERVATION LAWS}
\author{Sidney Bludman}
\email{sbludman@das.uchile.cl}
\homepage{http://www.das.uchile.cl/~sbludman}
\affiliation{Departamento de Astronom\'ia, Universidad de Chile, Santiago, Chile}
\author{Dallas C. Kennedy}
\email{dkennedy@mathworks.com}
\homepage{http://home.earthlink.net/~dckennedy}
\affiliation{The MathWorks, Inc., 3 Apple Hill Drive, Natick, Massachusetts, USA}
\date{\today}

\begin{abstract}
Scaling symmetries of the Euler-Lagrange equations are generally
not variational symmetries of the action and do not lead to
conservation laws.  Nevertheless, by an extension of Noether's theorem,
scaling symmetries lead to useful {\em nonconservation} laws,
which still reduce the Euler-Lagrange equations to first order in terms of scale invariants.
We illustrate scaling symmetry dynamically
and statically. Applied dynamically to systems of bodies
interacting via central forces, the nonconservation law is
Lagrange's identity, leading to generalized virial laws.
Applied to self-gravitating spheres in hydrostatic equilibrium,
the nonconservation law leads to well-known
properties of polytropes describing degenerate stars and chemically homogeneous nondegenerate stellar cores.
\end{abstract}

\maketitle
\tableofcontents

\section{SCALING SYMMETRY NOT GENERALLY A SYMMETRY OF THE ACTION}

Action principles dominate physical theories because they admit transformations among dynamical variables and
exhibit common structural analogies across different systems. If
these transformations are symmetries of the action, then by Noether's theorem, they give rise to {\em conservations
laws} that reduce the number of degrees of freedom.  This relationship of symmetries of the action (variational
symmetries) to conservation laws is central to Lagrangian dynamics.  Even if these transformations are not symmetries of
the action, they nonetheless lead to a useful {\em Noether's identity}.

Although equations of motion do not require Lagrangian expression, we apply Noether's identity to transformations that
are not symmetries; in particular, to scaling symmetry, which is not generally a symmetry of the action, but only of
the equations
of motion (Section~II).  Variational symmetries and generalized symmetries both reduce the equations of motion to first
order, but in different ways.

$\bullet$ Variational symmetries imply conservation laws, first integrals of the equations of motion.

$\bullet$ Scaling symmetry generally implies only a {\em nonconservation law}, which still reduces the equations of
motion to first order in scaling invariants.

Applied to dynamical systems of bodies interacting via inverse power-law potentials, these nonconservation laws are
Lagrange's formulae, or generalized virial theorems (Section~III). Applied to self-gravitating barotropic spheres
in hydrostatic
equilibrium (Section~IV), the nonconservation law leads directly to a first-order equation for homology invariants and
to well-known properties of polytropes and of homogeneous stellar cores (Section~V).  In this way,
these nonconservation laws illuminate the physical consequences of scaling symmetry.

The applications of continuous local symmetries of classical Lagrangians considered here are actually few.  We do not
consider applications to quantum field theories~\cite{Peskin}, involving the symmetry of the
vacuum state as well as the Lagrangian, which lead to important quantum anomalies and
topological symmetries, generated by topological charges.

\section{NOETHER'S THEOREM EXTENDED TO SCALING SYMMETRIES} %p2

\subsection{Noether's Identity Implies Both Conservation and Nonconservation Laws}

We consider a one-dimensional discrete dynamical system described by the Lagrangian density
$\mathcal{L}(t,q_i,\dot{q_i})$ and action $S=\int{\mathcal{L}(t,q_i,\dot{q_i})} dt$, where the dot designates the
partial derivative $\partial/\partial t$ with respect to the independent variable. Under the infinitesimal point
transformation
$\delta (t,q_i), \delta q_j (t,q_i)$ generated by $\delta t \cdot\partial/\partial t+
\delta q_i\cdot\partial/\partial q_i$,
the partial derivative and the Lagrangian transform locally as
\bea \delta \dot{q_i}=\frac{d \delta q_i}{d t}-\dot{q_i}\frac{d \delta t}{d t} \quad , \nonumber \\
\delta \mathcal{L}=[\delta t\cdot\partial/\partial t+
\delta q_i\cdot\partial/\partial q_i+\delta\dot{q_i}\cdot\partial/\partial\dot{q_i}]\mathcal{L}=
\dot{\mathcal{L}}\delta t+(\partial\mathcal{L}/\partial q_i)\delta q_i+(\partial\mathcal{L}/\partial\dot{q_i})
\Bigl[\frac{d \delta q_i}{d t}-\dot{q_i}\frac{d \delta t}{d t}\Bigr] \quad ,\eea
where the total derivative $d/dt\equiv\partial/\partial t+\dot{q_i}\cdot\partial/\partial q_i$. The Einstein summation
convention is assumed.

The canonical momentum and Hamiltonian
\be p_i(t,q_i,\dot{q_i}):=\partial\mathcal{L}/\partial\dot{q_i} \quad ,
\quad h(t,q_i,\dot{q_i}):=\dot{q_i}( \partial\mathcal{L}/\partial\dot{q_i} )-\mathcal{L} \ee
have total derivatives
\bea \frac{d p_i}{d t}=\partial\mathcal{L}/\partial q_i- \mathcal{D}_i \quad ,
\quad -\frac{dh}{dt}=\dot{ \mathcal{L} }+\mathcal{D}_i\cdot \dot{q_i}\quad , \nonumber \\
\frac{d}{dt} (p_i \delta q_i)=(\partial\mathcal{L}/\partial q_i -\mathcal{D}_i) \delta q_i +
  p_i\cdot \frac{d(\delta q_i)}{dt}\quad ,
\quad \frac{d (-h\delta t)}{dt}=(\dot{\mathcal{L}}+\mathcal{D}_i\cdot \dot{q_i}) \delta t -
h\cdot\frac{d(\delta t)}{dt}\quad ,\eea
in terms of the Euler-Lagrange variational derivative $\mathcal{D}_i:=
  \partial\mathcal{L}/\partial q_i-d(\partial\mathcal{L}/\partial\dot{q_i})/dt$.
The {\em Noether charge}
\be G:=\mathcal{L}\cdot\delta t+p_i\cdot(\delta q_i-\dot{q_i}\delta t)= -h\delta t+p_i\delta q_i\ee
has total derivative
\be \frac{dG}{dt}\equiv\delta\mathcal{L}+\mathcal{L}\cdot\frac{d(\delta t)}{dt}-
\mathcal{D}_i\cdot(\delta q_i-\dot{q_i}\delta t)=\bar{\delta}\mathcal{L}-
  \mathcal{D}_i\cdot(\delta q_i-\dot{q_i}\delta t)\quad , \label{eq:Noetheridentity} \ee where
$\bar{\delta}\mathcal{L}:=\delta\mathcal{L}+\mathcal{L}\cdot (d\delta t/dt)$
is the change in Lagrangian at a fixed point. Different Lagrangians, leading to
the same equations of motion, define different Noether charges and nonconservation laws.

The variation in action between fixed times and end points is
\be \delta S_{12}=\int_2^1 dt\ \delta\mathcal{L}=\int_2^1 dt\ \Bigl[\frac{dG}{dt}-
  \mathcal{L}\cdot\frac{d(\delta t)}{dt}+\mathcal{D}_i\cdot(\delta q_i-\dot{q_i}\delta t)\Bigr]=
  G(1)-G(2)+\int_2^1 dt\ \Bigl[\delta q_i\cdot\mathcal{D}_i+\delta t\cdot \Bigl(\frac{dh}{dt}+
  \frac{\partial\mathcal{L}}{\partial t}\Bigr) \Bigr]\quad , \ee
after integrating the term in $d(\delta t)/dt$ by parts.
The action principle asserts that this variation vanishes for independent variations $\delta q_i, \delta t$ that vanish
at the end points.  It implies the Euler-Lagrange equations and $dh/dt=-\partial\mathcal{L}/\partial t$. On-shell, where
the equations of motion $\mathcal{D}_i=0$ hold,
\bea \delta S_{12}=\int_2^1{ \bar{\delta}\mathcal{L}}\ dt=G(1)-G(2) \\
\fbox{$ \displaystyle \bar{\delta}\mathcal{L}=\frac{dG}{dt} $}\quad . \eea
The last result is {\em Noether's equation}, identifying the total derivative of the Noether charge with the change in
Lagrangian that generates it. By expressing the equations of motion in divergence-like form, it has an important
consequence: a {\em conservation law}, if $G$ is a variational symmetry; a {\em nonconservation law} otherwise.

\subsection{Transformations That Are Not Symmetries Still Lead to Useful Nonconservation Laws} %p3

To the nonrelativistic central-force system
\be h(\mathbf{r},\mathbf{p})=K+V(r):=\mathbf{p}^2/2m +V(r) \quad , \quad p^2=p_r ^2+l^2/r^2
  \quad ,\label{eq:centralforce}\ee
apply the static radial dilatation~(A) $\delta\mathbf{r}=\mathbf{r}$ and radial translation~(B)
$\delta\mathbf{r}=\mathbf{r}_1:=\mathbf{r}/r$:
\begin{enumerate}
\item[(A):] $G:=\mathbf{p}\cdot\delta\mathbf{r}=\mathbf{r}\cdot\mathbf{p}$:
$\delta\mathbf{r}=\mathbf{r} \quad ,\quad \delta\mathbf{p}=-\mathbf{p}\quad ,\quad\qquad\qquad\qquad
  \delta h=-2K+r(dV/dr)$
\item[(B):] $G:=\mathbf{p}\cdot\delta\mathbf{r}= p_r$:
~~$\delta\mathbf{r}=\mathbf{r}_1\quad ,\quad\delta\mathbf{p}=-\mathbf{p}_t/r=\mathbf{r}\times\mathbf{l}/r^3
  \quad ,\quad\delta h=-l^2/mr^3 + dV/dr\quad ,$
\end{enumerate}
where
\be \mathbf{p}_r=(\mathbf{r}\cdot\mathbf{p})\mathbf{r}/r^2\quad , \quad \mathbf{p}_t=\mathbf{p}-\mathbf{p}_r =-\mathbf{r}\times\mathbf{l}/r^2
\quad , \quad \mathbf{l}:=\mathbf{r}\times \mathbf{p} \ee
are the radial and transverse linear momenta, and the angular momentum, respectively~\cite{Landau}.

Because both these radial transformations are static, $G=\mathbf{p}\cdot\delta\mathbf{r}$, $dG/dt=\delta\mathcal{L}=
-\delta h$, the two nonconservation laws are
\begin{enumerate}
\item[(A):] $dG/dt=d(\mathbf{r}\cdot\mathbf{p})/dt=2K-r(dV/dr)$
\item[(B):] $dG/dt=dp_r/dt=l^2/m r^3-dV/dr$\quad .
\end{enumerate}
Except for circular orbits, neither of these transformations is a symmetry.  Nonetheless, each of these
nonconservation laws expresses important consequences of the equations of motion, in any
central-force system~(\ref{eq:centralforce}).
\benum
\item[(A):] Defining the {\em virial} $\mathbf{p}\cdot\mathbf{r}:=A$, the $\mathbf{r}\cdot\mathbf{p}$ nonconservation
law is Lagrange's formula $\dot{\mathbf{p}}=-\nabla V$, which preceded Clausius by almost a century~\cite{Doughty}.
In the form $dA/dt=2K+\sum\mathbf{r}_i\cdot\mathbf{F}_i$, the law still holds for a system of bodies,
even if the central forces $\dot{\mathbf{p}_i}=\mathbf{F}_i$ do not derive from a potential.
\item[(B):] The $p_r$ nonconservation law is the radial equation of motion.
\eenum

Both these nonconservation laws express the equations of motion and do not depend on scaling symmetry.  But,
if the potential is homogeneous in $r$, so that $r(dV/dr)$=$-nV$, the system is scaling symmetric. In any bounded
ergodic system, the time averages $\langle dA/dt\rangle = \langle dp_r/dt \rangle$ vanish, so that
\begin{description}
  \item[(A):] $2\langle K\rangle =-n\langle V\rangle$
  \item[(B):] $(l^2/m)\langle 1/r^3\rangle =-n\langle V/r\rangle\quad , \quad l \neq 0$\quad .
\end{description}
For $n=1$, (B)~is useful for relativistic corrections to noncircular hydrogenic orbits.  (A)~is the usual virial law.

In nondegenerate perfect gases, equipartition makes the internal gas kinetic energy
$K=\frac{3}{2}\int P dV$=$\int\varepsilon dm$, where the internal energy density
$\varepsilon=\frac{3}{2}(P/\rho)=\frac{3}{2}\mathcal{R}T/\mu$ for a gas of molecular weight $\mu$. The Coulombic
virial theorem $2 \langle K\rangle =-\langle V\rangle$ then determines the
averaged specific temperature $\langle T/\mu\rangle $ and leads to important applications in classical kinetic theory
and in stellar structure.

\subsection{Variational Symmetries Imply Conservation Laws} %%p4

The most general and important applications of Noether's identity are to variational symmetries and to dynamic scaling
symmetries of the equations of motion, which preserve the stationary action principle $\delta S_{12}=0$ and reduce
the equations of motion to first order in different ways.

Variational symmetries preserve the action $\delta S_{12}=0$ because $\bar{\delta}\mathcal{L}=0$ or $dB/dt$, the
total derivative of some gauge term $B(t,q)$.  Noether's identity $(d/dt)(G-B) \equiv -
\mathcal{D}_i\cdot({\delta q_i}-\dot{q_i}\delta t)$ conserves $G - B$ on-shell, when the equations of motion hold.  This
original version of Noether's theorem, identifying conservation laws with variational symmetries, has two familiar
applications.
 \begin{description}
   \item[Point symmetries lead to integration by quadratures:] Any central-force system~(\ref{eq:centralforce}) is symmetric under time
   translations and spatial rotations, leading to conservation of energy $E$ and angular momentum $l$:
       \be E:=(\dot{r}^2+r^2\dot{\theta}^2)/2+V(r)\quad , \quad l:= m r^2 \dot{\theta}\quad .\ee
       Since $r\dot{\theta}/\dot{r}=p_t/p_r=\sqrt{r^2 p^2/l^2-1}$, quadrature leads to the first-order orbit and time equations
       \be \theta(r)=\theta_0 + \int dr/r \sqrt{2 m r^2 [E-V(r)]/l^2-1}\quad ,
       \quad t=t_0+\int r dr/\sqrt{2 r^2 [E-V(r)]-(l/m)^2}\quad .\ee
       In the Newtonian case $V(r)=-G M/r$, the integrals reduce to elementary functions and the orbits are
       conic sections
       \be r(\theta)=p/[1-\epsilon\sin (\theta-\theta_0)] \ee
       of eccentricity $\epsilon^2$=$1+2 E(l/G m M)^2$, where $p:=(l/m)^2/GM$~\cite{Landau}.
   \item[Conservation laws including any gauge terms:] A variational symmetry in which the Noether charge is not
conserved obtains in
the many-body system of particles with interparticle forces that depend only on the relative separations
$\mathbf{q}_i -\mathbf{q}_j$ and relative velocities $\dot{\mathbf{q}}_i -\dot{\mathbf{q}}_j$.  This system admits the
{\em infinitesimal boost transformations}
\be\delta \mathbf{q}_i = \delta\mathbf{v}\cdot t\quad ,\quad \delta t=0\quad ,\quad \delta V=0\quad ,
\quad \delta K=\delta\mathcal{L}=\mathbf{P}\cdot\delta v\quad ,\ee
where $M,\mathbf{P},K$ are the total mass, momentum, and kinetic energy. The Noether charge $G=(\mathbf{P} \cdot
\mathbf{v}) t$ is not conserved, but Noether's equation gives the conservation law $(\mathbf{P}-M\dot{\mathbf{R}}) \cdot
\delta \mathbf{v}=0$, or  $M\dot{\mathbf{R}}=
\mathbf{P}$, for arbitrary infinitesimal  $\delta\mathbf{v}$. Boosts change the total momentum $\mathbf{P}$, but the
center-of-mass moves with velocity $\dot{\mathbf{R}}$.  This familiar center-of-mass theorem follows directly from
boost symmetry, irrespective of the internal forces. It is paradigmatic for distinguishing between the effects of
internal and external forces on many-body system.
\end{description}

The converse of Noether's theorem is that conservation laws imply invariance of the Lagrangian up to a possible gauge
term.  For example, the conservation of the relativistic momentum and energy implies Lorentz invariance of the
Lagrangian.

\subsection{Scaling Symmetry Implies a Special Nonconservation Law}  %%p6

In a many-body system with individual coordinates $\mathbf{r}_i$, the dynamical scale transformation
\be \delta t=\beta\cdot t\quad , \quad \delta \mathbf{r}_i =\mathbf{r}_i\quad ,\quad \delta (\partial/\partial t)=
-\beta\cdot (\partial/\partial t)\quad ,\quad \delta\dot{\mathbf{r}_i}=(1-\beta)\cdot\dot{\mathbf{r}_i}\quad ,
  \quad \delta (r^{\beta}/t)=0
\label{eq:dynscaletrans} \ee
is generated by the Noether charge
\be G:=-\beta\cdot h t +A\quad , \ee
where $A:=\sum{\mathbf{p}_i \cdot \mathbf{r}_i}$ is the virial. Dynamical scaling is a symmetry of the equations of
motion (but not of $\mathcal{L}$), if the pairwise potential energies are inverse powers $V_{ij}\sim
|\mathbf{r}_i-\mathbf{r}_j|^{-n}$ of the interparticle distances, the potentials are homogeneous in their coordinates
$r(dV_{ij}/dr) = -n V_{ij}, \delta V_{ij}=-n V_{ij}$ and
$\beta =1+n/2$, so that all distances scale as $r_i\sim t^{1/\beta}=t^{n/(2+n)}$~\cite{Landau}.

Scaling symmetry makes the Lagrangian a homogeneous function of its arguments, a scalar density of some weight
$-2\tilde{\omega}$, so that $\delta\mathcal{L}=-2\tilde{\omega}\mathcal{L},
~\delta S_{12}=\tilde{\sigma} S_{12}$, where $\tilde{\sigma}:=1-2 \tilde{\omega}$.
Noether's identity then implies the special on-shell {\em scaling nonconservation law}
\be \frac{d G}{d t}=(1-2\tilde{\omega})\mathcal{L}=\tilde{\sigma}\mathcal{L}\quad ,\ee                  %%Eq. (14)
which reduces to a conservation law only asymptotically,
wherever $\tilde{\sigma}\mathcal{L}$ is small.
This asymptotic conservation law then allows approximate integration of the equations of motion, in certain limits.

In the next section, we consider energy-conserving mechanical systems
$\mathcal{L}$=$\mathcal{L}(\mathbf{r}_i,\dot{\mathbf{r}}_i)$, for which
the dynamic scaling nonconservation law is a generalized virial law.  In Section~IV, we consider the
spherical hydrostatics of barotropic fluids, for which the Lagrangian $\mathcal{L}(r,H,H')$ depends explicitly on the
independent variable $r$.  Instead of a first integral, both these examples illustrate a first-order differential
equation among homology invariants~\cite{Olver,Blumen}, linearly relating the ``energy function'' $h=K+V$
to $\mathcal{L}=K-V$, or the ``kinetic'' term $K$ to the ``potential'' term $V$.

\begin{table}[b] %%%%Table I p5
\caption {Period-Amplitude Relations and Virial Theorems for Inverse Power-Law Potentials $V \sim 1/r^n$}
%\begin{ruledtabular}
\begin{tabular}{|l|l|l|l|}
\hline\hline
$n$&System&Period-amplitude relation $t\sim r^{1+n/2}$&Virial theorem\\
\hline\hline
-2 & isotropic harmonic oscillator  & period independent of amplitude       & $\langle K\rangle =\langle V\rangle$ \\
-1 & uniform gravitational field    & falling from rest, e.g., $z=gt^2 /2$  & $\langle K\rangle =\langle V\rangle /2$ \\
0  & free particles                 & constant velocity $r\sim t $          & $\langle K\rangle =0$                \\
1  & Newtonian potential            & Kepler's Third Law $t^2\sim r^3$      & $\langle K\rangle =-\langle V\rangle /2$ \\
2  & inverse cube force             & $t \sim r^2 $                         & $\langle K\rangle =-\langle V\rangle $ \\
\hline\hline
\end{tabular}
%\end{ruledtabular}
\end{table}

\section{DYNAMICAL NONCONSERVATION LAWS FOLLOWING FROM SCALING SYMMETRY} %%IIIp5

\subsection{Mechanical Nonconservation Laws Are Generalized Lagrange Identities}

Consider a nonrelativistic system of particles with coordinates $\mathbf{r}_i$, momenta $\mathbf{p}_i$, interacting by
pairwise static potential energies $V_{ij}$. The dynamical scale transformation~(\ref{eq:dynscaletrans}) generates the
infinitesimal changes
 \be \delta K=2(1-\beta) K\quad ,\quad \delta V =-(1-\beta)\mathbf{r}\cdot \nabla V\quad , \quad \delta\dot{A}=
 (1-\beta)\dot{A}\quad .\ee
If the pairwise potential energies are inverse powers $V_{ij}\sim |\mathbf{r}_i-\mathbf{r}_j|^{-n}$ of the interparticle
distances, the potentials are homogeneous in their coordinates $r_{ij}(dV_{ij}/dr_{ij}) = -n V_{ij},
\delta V_{ij}=-n V_{ij}$. Provided $n\equiv 2(\beta-1)$, $\beta\equiv 1+n/2$, the Lagrangian density scales as
\be \delta\mathcal{L}=-n\mathcal{L}\quad ,\quad \bar{\delta}\mathcal{L}=\delta\mathcal{L}+\beta\mathcal{L}=
(1-n/2)\mathcal{L}\quad . \ee
Because energy is conserved,
$d G/d t=-\beta h+\dot{A}$, so that the scaling symmetry nonconservation law
\be \dot{A}=(1+n/2)h+(1-n/2)\mathcal{L}=2 K+ n V \ee
relates the nonrelativistic kinetic energy $K$ and power-law potential $V$ to the time derivative of the virial $A$.

For periodic or long-time averages in bounded ergodic systems, we have $\langle\dot{A}\rangle$ = 0, and the
{\em virial theorem} $2\langle K\rangle$ = $-n\langle V\rangle $.  Table~I tabulates these generalized virial
theorems and period-amplitude relations for orbits in the five important inverse-power-law potentials $n=-2,-1,0, 1, 2$.
Only for $n=2$ does dynamical scaling reduce to a variational symmetry, so that the Noether charge $G=-2(K+V)t+A$ is
conserved. For potentials more singular than $1/r^2$, there are no bound states.

\subsection{Scaling Nonconservation Law in Classical Electrodynamics}%IIIB p6

Noether's identity applies to continuous Lagrangian systems (fields) as well as discrete systems. In this case,
$\mathbf{r},t$ are independent variables.  If $\mathbf{f},~\mathbf{G}=
\mathbf{E}\times\mathbf{B}/4\pi c,~\mathbf{\mathcal T},~U$
are respectively the electromagnetic force density, momentum density, momentum flux tensor, and energy density,
then momentum balance reads
\be \partial \mathbf{G}/\partial t +\nabla \cdot \mathbf{\mathcal T}+\mathbf{f} =0\quad . \ee
From this follows an electromagnetic analogue of the mechanical Lagrange's identity:
\be \partial(\mathbf{r}\cdot \mathbf{G})/\partial t+\nabla\cdot(\mathbf{\mathcal T}\cdot\mathbf{r})-
U+\mathbf{r}\cdot\mathbf{f}=0 \quad .\ee
When time-averaged, this becomes an electromagnetic virial theorem~\cite{Schwinger}.

\subsection{Scaling Nonconservation Law in Classical Conformal Field Theory}

In any relativistic field theory, space-time scaling (dilatation) symmetry leads to the familiar nonconservation law
\be \frac{\partial G^{\mu}}{\partial x^{\mu}}=\Theta^{\mu}_{\mu}\quad ,\ee
where $\partial /\partial x_{\mu}$ is the four-dimensional divergence, $G^{\mu}$ is the dilatation current, and
$\Theta^{\mu}_{\mu}$ is the trace of the energy-momentum tensor~\cite{CallanColeman,Coleman,Peskin}.  The dilatation
charge is conserved when this trace vanishes, implying a conformal symmetry.

The most familiar example of conformal symmetry is Laplace's equation in $n$ spatial dimensions. In two dimensions,
conformal symmetry implies the Cauchy-Riemann equations, so that any analytic function is a solution of
Laplace's equation.  In higher dimensions, conformal symmetry implies the conservation laws associated with
translations, rotations, dilatations, and spatial inversions.  The pure electromagnetic field is conformally invariant.

These dynamical systems illustrate how Noether's identity leads to useful and often familiar nonconservation laws, even
when scaling symmetry is broken. The remainder of this paper considers the hydrostatic equilibrium of gaseous spheres,
where the independent variable is the radial coordinate $r$ and the variational principle is that of minimum energy, in
place of least action.

\section{SCALE-INVARIANT BAROTROPIC SPHERES} %%p6

\begin{figure}[t] %%%Fig. 1 p7
\includegraphics[scale=0.70]{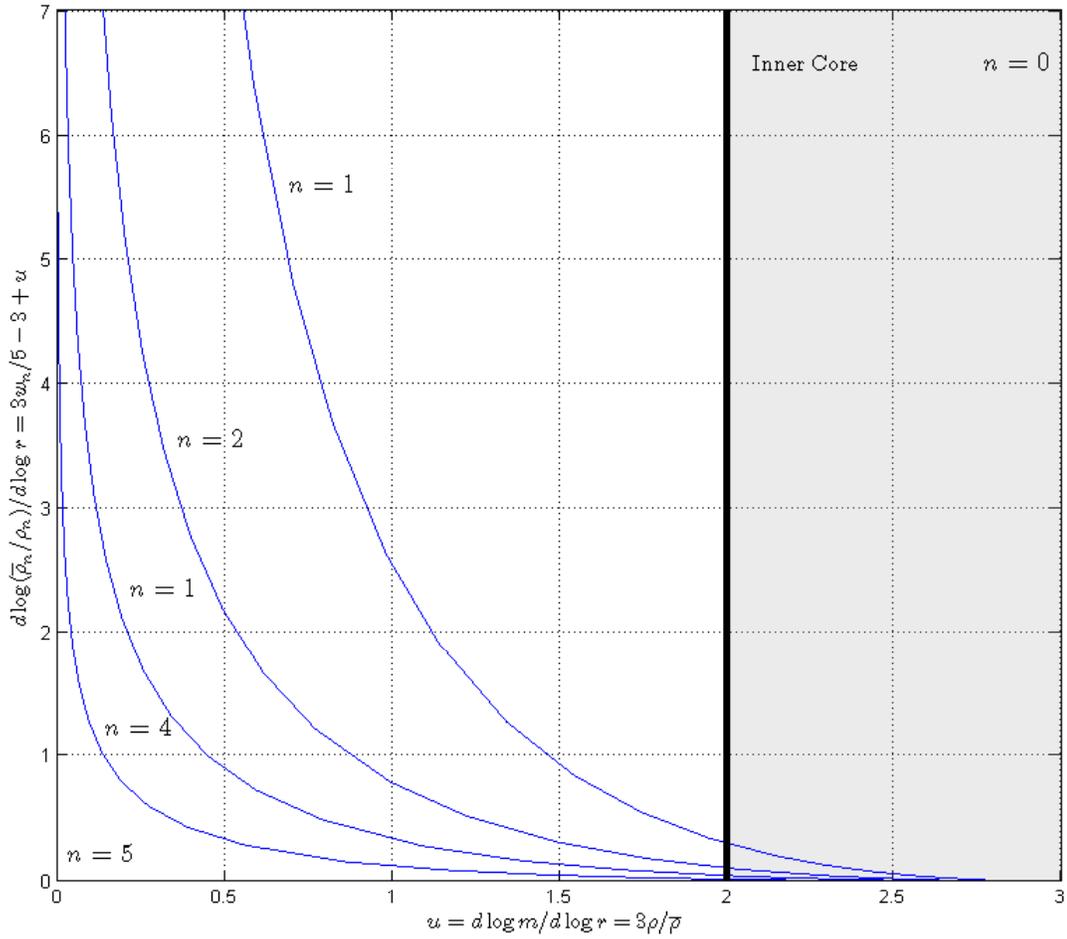}
\caption
{Effect of the outer boundary on the Emden polytrope density profiles.  For the softest equation of state $n=5$,
the stellar radius is infinite, $w_5=(5/3)(3-u)$ everywhere, and $\bar{\rho}=\rho_c ^{2/5} \rho ^{3/5}$. For stiffer
equations of state $n<5$, $w_n\approx w_5$ inside the core $u<2$, but increases as the finite radius is approached
$u \rightarrow 0$.}
\end{figure}

The structure of luminous stars depends on the coupling between hydrostatic and thermal structure through an equation
of state $P=P(\rho, T)$, which generally depends on the local temperature and chemical composition.  Ignoring chemical
evolution, the matter entropy is locally conserved, so that in the steady state, stars are in both local thermal and
chemical equilibrium.  To treat the hydrostatic equilibrium
independently of heat flow, we consider only stars in which the thermal structure is specified independently, so that
the local equation of state is {\em barotropic} $P=P(\rho)$ and $dP/\rho=dH$ in terms of the specific enthalpy
$H(r)=E+P/\rho$.  This restriction to barotropic stars makes the density $\rho(r)$, specific internal energy $E(r)$,
specific enthalpy $H(r)$ = $E+P/\rho$, and {\em thermal gradient} $\nabla (r):= d \log{T}/d \log{P}$ implicit functions
of the gravitational potential $V(r)$.  The assumption of the independence of
the thermal structure is justified as a good approximation if, as is usually the case, the thermal (Kelvin-Helmholtz)
relaxation time of the whole star is much longer than its hydrostatic equilibration time.

The hydrostatic structure of barotropes depends only on two first-order equations, mass continuity and
of pressure equilibrium,
\be d m/d r=4\pi r^2 \rho \quad, \quad -dP/dr=G\rho m/r^2\quad ,\quad \label{eq:masspressbalance}\ee
or
\be \frac{1}{r^2}\frac{d}{dr}\Bigl(\frac{r^2}{\rho}\frac{dP}{dr}\Bigr) = -4\pi G\rho\quad ,\quad
(r^2 H')'+4\pi G r^2 \rho(H)=0 \label{eq:hydrostaticEL-1}\quad ,
\ee
where $':=d/dr$. In terms of the specific gravitational force $d V/dr=g:=Gm/r^2$, the equation of
hydrostatic equilibrium reads
\be d (H+V)/d r=0 \quad .\ee

These structural equations are the Euler-Lagrange equations of the Lagrangian
\be \mathcal{L}(r,H,H')=4\pi r^2[-(H')^2/8\pi G+P(H)]\quad ,~\label{eq:hydrostaticLagrange-2} \ee
derived from a minimal energy variational principles in Appendix A. However, the following consequences of assuming scale invariance do not depend explicitly on a Lagrangian formulation or on Noether's identity.

\subsection{In a Simple Ideal Gas, Scale Invariance Requires a Constant Entropy Gradient } %IV.A p7

{\em Polytropes} are barotropic spheres with constant polytropic exponent $1+1/n:=d\log{P}/d\log{\rho}$ and
polytropic index $n=d\log{\rho}/d\log{H}$. The pressure, specific energy, specific enthalpy, enthalpy gradient and
central pressure at any point are
\be P/P_c=(\rho /\rho_c)^{1+1/n}\quad ,\quad E=n(P/\rho)\quad ,\quad H=(n+1)(P/\rho)\quad ,\quad
d\log{H}/d\log{P}=1/(n+1)\quad .\ee
Polytropic mechanical structure does not fix the thermal structure, which depends on the heat transport mechanism.

In a simple ideal gas, the equation of state, specific internal energy, specific enthalpy and adiabatic exponent are
\be P/\rho=\frac{\mathcal{R}}{\mu} T\quad ,\quad E=C_V T\quad ,\quad H=C_P T\quad ,
\quad (d\log{P}/d\log{\rho}_S\quad , \ee
where $\mathcal{R}$ is the universal gas constant, $\mu$ is the molecular weight, and $C_P,~C_V$ are the specific heats
at constant pressure and density.  Following the law of energy conservation (Appendix~B), the specific entropy and
thermal gradient of a simple ideal gas are
\be ~dS=C_V d\log{P}-C_P d\log{\rho}\quad ,\quad S=C_V\log{(P/\rho^{ \frac{C_P}{C_V}})}\quad ,
\quad \nabla=d\log{H}/d\log{P}\quad .\ee

Bound in a polytropic structure of index $n$, an ideal gas has constant thermal gradient, gravithermal specific heat,
and entropy-pressure gradient:
\be \nabla=1/(n+1)\quad ,\quad C^{*}=C_P(1-\nabla_{ad}/\nabla)\quad ,
\quad dS/d\log{P}=C_P (\nabla-\nabla_{ad})\quad .\ee
The radial entropy gradient
\be dS/d\log{r}=C_P (\nabla_{ad}-\nabla)\cdot v_n \quad , \ee
is proportional to the homology invariant $v_n$ and positive (zero) when the thermal gradient is subadiabatic (zero).
(See Section~IV.B for more about $v_n$.)
In convective equilibrium, any polytrope has constant entropy $S$. In radiative equilibrium, a simple ideal gas polytrope has constant temperature and entropy gradient $d S/d \log{P}$.

The thermal gradient is nearly constant and the hydrostatic structure nearly polytropic in zero-temperature (degenerate)
stars and in chemically homogeneous stars starting out on the hydrogen-burning, zero-age Main Sequence (ZAMS):
\begin{description}
  \item[White dwarfs and neutron stars:] nonrelativistic and extreme relativistic degenerate stars; polytropes of
  index $n$=3/2 and 3, respectively.
  \item[ZAMS stars in convective equilibrium:] with vanishing gravithermal specific heat $C^{*}$ and uniform entropy
      density.  These are $n$=3/2 polytropes.
  \item[ZAMS stars in radiative equilibrium:] With uniform energy generation and Kramers opacity, stable polytropes of
      $n>3/2$. At zero age, our Sun was a chemically homogeneous star of mean molecular weight
  $\mu=0.61$, well-approximated by the Eddington standard model ($n$=3) throughout its radiative zone, which contained
  99.4\% of its mass. Because energy generation was centrally concentrated, our ZAMS Sun was even better approximated by
  a slightly less standard $n$=2.796 polytrope~\cite{BludKen}.
  \item[] Even better nonpolytropic fits, to the $M-R$ relation $R\sim M^{\xi}$ observed in
young ZAMS stars are obtained by including nonuniform energy transport and corrections to Kramers opacity:
radiative transport in the $pp$-burning lower main sequence $0.11<M/M_\odot <1.2$ gives $\xi=0.57$;
convective transport in the CNO-burning upper main sequence $2 <M/M_\odot <20$ gives $\xi=0.8$~\cite{Kippen,Hansen}.
\end{description}

Chemically inhomogeneous stars and the photospheres of luminous stars cannot be polytropic. Because our present Sun is
chemically evolved and has a convective envelope, it is far from being polytropic: its polytropic fit, with
index $n=3.26$, is poor~\cite{BludKen}.

\begin{figure}[t] %Fig. 2 p9
\includegraphics[scale=0.70]{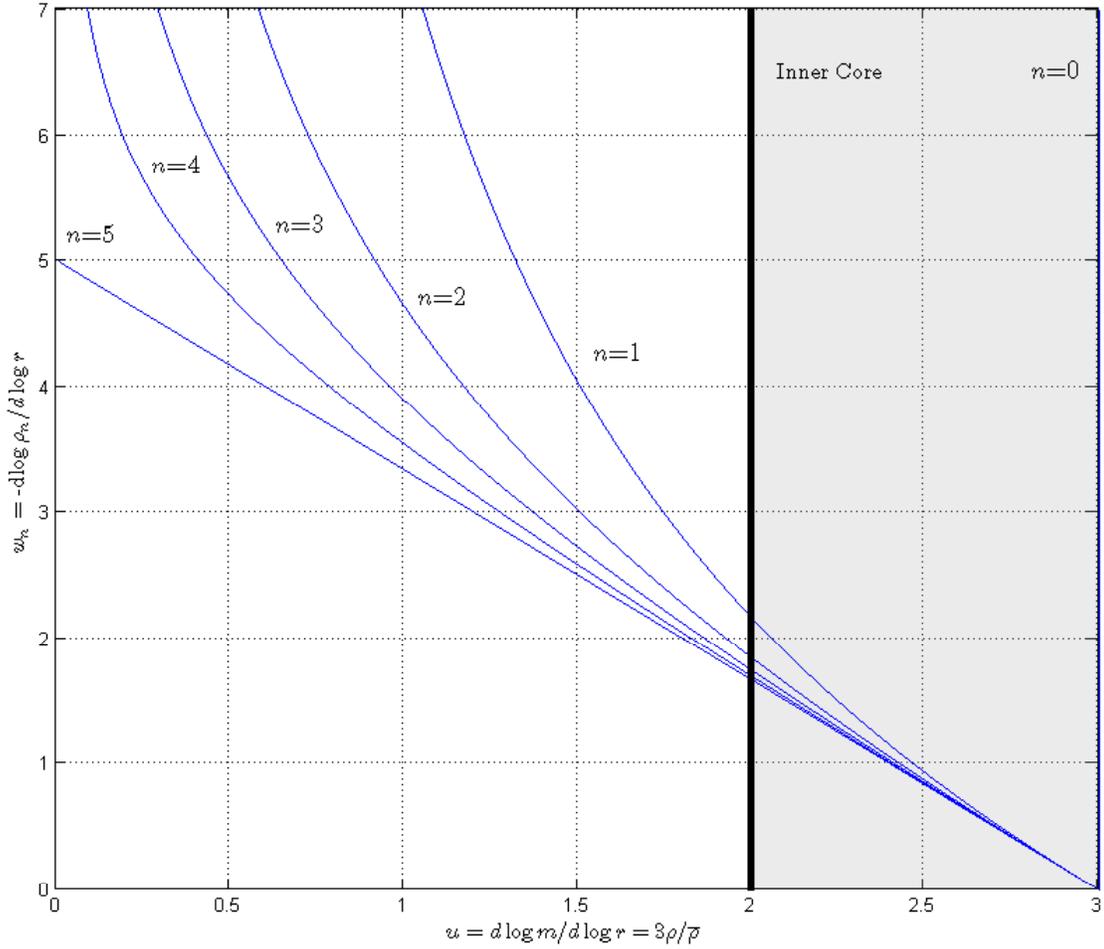}
\caption{Dilution of polytrope density profiles in the envelope as the boundary is approached ($u\rightarrow 0$). All
solutions approach the same density structure $w_n(z)\rightarrow w_5=(5/3)(3-u)$ at the center ($u\rightarrow 3$), but
differ outside the core. As the boundary is approached, the mass $m\rightarrow M$, and the density scale height
$r/w_n\rightarrow 0$.}
\end{figure}

\subsection{Polytropic Structure Implies a First-Order Equation in Scaling Invariants} %%IV.B p8

Following Chandrasekhar~\cite{Chandra}, we define homology variables
\be u:=d \log{m}/d\log{r}=3\rho/\bar{\rho}\quad ,\quad v:=-d\log{(P/\rho)}/d\log{r}\quad ,
\quad w:=-d\log{\rho}/d\log{r}=n(r)\cdot v\quad , \ee
where $\bar{\rho}=3 m/4\pi r^3$ is the average mass density interior to radius $r$ and
$n(r):=d\log{\rho}/d\log{(P/\rho)}$. The central boundary condition is
\be u(0)=3\quad ,\quad v(0)=0\quad ,\quad (d v/du)_0=-5/3n\quad .\ee
The mass continuity and hydrostatic equilibrium equations~(\ref{eq:masspressbalance}) become
\be d\log{u}/d\log{r}=3-u-n(r) v\quad ,
\quad d\log{v}/d\log{r}=u-1+v-d\log{[1+n(r)]}/d\log{r}\quad ,\label{eq:autonomous} \ee
are autonomous only when the index $n(r)$ is constant.

In polytropes, the constant index $n$ and gradient $\nabla=1/(n+1)$ imply
\be du/d\log{r}=u(3-u-n v_n)\quad ,\quad dv_n/d\log{r}=v_n(u-1+v_n)\quad ,\ee
which are both autonomous and can be written as the {\em characteristic equations}
\be \frac{d\log{v_n}}{u-1+v_n}=\frac{d\log{u}}{3-u -n v_n}=d\log{r} \label{eq:chareqns} \ee
for the phase variables $u,~v_n$. Deferring the second equality for $r=r[v_n(u)]$ to the next section,
we now solve the first-order Abel equation
 \be \frac{dv_n}{du}=\frac{v_n(u-1+v_n)}{u(3-u-n v_n)}\quad ,\ee
subject to the central boundary condition $u(0)=3, v_n(0)=0$ for
{\em regular (Emden) polytropes} $v_n(u)$~\footnote{
This first-order equation is important elsewhere in mathematical physics. In population dynamics, with $\log{r}$
replaced by
time $t$, it becomes the Lotka-Volterra equation for predator/prey evolution~\cite{Boyce,JordonSmith}. The $u v_n$
cross-terms lead to growth of the predator $v_n$ at the expense of the prey $u$, so that a population that is
exclusively prey initially ($v_n=0,u=3$) is ultimately devoured $u\rightarrow 0$. For the weakest predator/prey
interaction ($n=5$), the predator takes an infinite time to reach only the finite value $v_n=1$.
For stronger predator/prey interaction ($n<5$), the predator grows infinitely $v_n\rightarrow\infty$ in a finite time.}.

The infinitesimal scale transformation
\be \delta r=r\quad ,\quad \delta H=-\tilde{\omega}H\quad ,\quad \delta H'=-(1+\tilde{\omega}) H' \ee
is generated by the Noether charge
\be G_n:=\Bigl[r \Bigl( \frac{H'^2}{2}+\frac{H^{n+1}}{n+1}\Bigr)+\tilde{\omega}\theta \theta' \Bigr] r^2\quad ,\ee
whose radial derivative
\be dG_n/d r \equiv \delta\mathcal{L}+\mathcal{L}+\mathcal{D}_r\cdot \tilde{\omega}\cdot d(Hr)/dr \quad , \ee
obeys the scaling nonconservation law
\be dG_n/d r=\tilde{\sigma}\mathcal{L}\quad ,\quad \tilde{\sigma}:=1-2 \tilde{\omega}\quad ,
\label{eq:scalingnonconslaw}\ee
wherever the Euler-Lagrange equation (25) is satisfied.
The structural equation is scale-invariant if and only if
\be n\tilde{\omega} = 2+\tilde{\omega}\quad ,
\quad 2(1+\tilde{\omega}) = (n+1)\tilde{\omega}\quad ,\quad \tilde{\omega} \equiv (n-5)/(n-1)\quad , \ee
so that the Lagrangian~(27) is homogeneous of degree $-2\tilde{\omega}$ and
$\delta\mathcal{L}=-2\tilde{\omega}\mathcal{L}$.
The scaling nonconservation law~(\ref{eq:scalingnonconslaw})
then connects the gravitational and internal energy densities,
just as the point-mechanics Lagrange identity connected the potential and kinetic energies.

For polytropes, we introduce dimensionless units
\be \xi:=r/\alpha\quad ,\quad \theta_n:=H/H_c=(\rho/\rho_c)^{1/n}\quad ,\ee
and the dimensional constant
\be \alpha ^2:=\frac{(n+1)}{4\pi G}K\rho_c ^{1/n-1}=(n+1)/4\pi G \cdot (P_c/\rho_c^2)\quad ,\ee
where $\rho_c$ is the central density and
\be P_c/\rho_c:= K\rho_c ^{1/n}=\theta_n^{1+n}\quad ,\quad H_c:=(n+1)P_c/\rho_c \equiv (n+1) K\rho^{1/n}\quad .\ee
The included mass, mass density, average mass density, and gravitational acceleration are
\bea
m=4\pi\rho_c\alpha^3\cdot(-\xi^2\theta_n ')\quad ,
\quad \rho=\rho_c\cdot\theta_n ^n\quad ,
\quad \bar{\rho}=\rho_c\cdot(-3\theta_n '/\xi)\quad ,\quad
g=4\pi\rho_c \alpha^2 (-\theta_n ')\quad .\eea
The Euler-Lagrange equation~(\ref{eq:hydrostaticEL-1}), combining mass continuity and hydrostatic equilibrium, takes the
dimensionless {\em Lane-Emden} form
\be \frac{d}{d\xi}\Bigl( \xi^2 \frac{d\theta_n}{d\xi}\Bigr) + \xi^2\theta_n ^n = 0\quad .\label{eq:lane-emden-1}\ee

In terms of $\theta,~\theta '$, the homology variables~\cite{Chandra}
\bea u:=d\log{m}/d\log{r}=3\rho(r)/\bar{\rho}=-\xi \theta^n_n/\theta_n ' \quad ,
\quad v_n:=-d\log{(P/\rho)}/d\log{r}= -\xi\theta_n'/\theta_n \quad ,
\quad \bar{\rho}/\rho_c=-3 \theta_n '/\xi^2 \quad , \nonumber \\
u/v_n=\theta_n^{n+1}/\theta_n '^2\quad ,\quad uv_n=\xi^2\theta_n^{n-1}=(\xi^{\tilde{\omega}}\theta_n)^{n-1}\quad ,
\quad \omega_n:=\xi^{\tilde{\omega}+1}(-\theta_n ')=(uv_n^n)^{\tilde{\omega}/2} \quad .\eea
Hereafter $':=d/d\xi$, and polytropes of different index are distinguished by the subscript $n$ attached to
different homology variables.

Extracting the dimensional constant $\mathcal{C}:=-H_c^2/G:=-[(n+1)K\rho_c^{1/n}]^2/G$ and suppressing the subscript
$n$ on $\theta$, the Lagrangian, Hamiltonian, and Noether charge are
\bea \mathcal{L}/\mathcal{C}=\xi^2[\theta '^2 /2-\theta^{n+1}/(n+1)]\quad ,\nonumber \\
\mathcal{H}/\mathcal{C}=\xi^2 [\theta'^2/2+\theta^{n+1}/(n+1)]\quad ,\nonumber \\
G_n/\mathcal{C}=\xi^2\Bigl[ \xi \Bigl( \frac{\theta'^2}{2}+\frac{\theta^{n+1}}{n+1}\Bigr)+
  \tilde{\omega}\theta \theta' \Bigr]\quad .\eea
The scaling nonconservation law~(\ref{eq:scalingnonconslaw})
\be \frac{d}{d\xi} \Bigl\{ \xi^2 \cdot \Bigl[ \xi\Bigl(\frac{\theta'^2}{2}+\frac{\theta^{n+1}}{n+1}\Bigr)+
  \tilde{\omega}\theta \theta' \Bigr] \Bigr\}=\tilde{\sigma}\xi^2
  \Bigl( \frac{\theta'^2}{2}-\frac{\theta^{n+1}}{n+1}\Bigr) \ee
is equivalent to the Lane-Emden equation~(\ref{eq:lane-emden-1}). It describes the evolving ratio between local
internal and (negative) gravitational energy densities
\be \frac{\theta^{n+1}/(n+1)}{\theta'^2 /2} = \frac{2}{n+1} \frac{u}{v_n}\quad ,\label{eq:energyratio}\ee
as the local energy density changes from entirely internal at the center, to entirely gravitational at
the stellar surface.

\subsection{Scaling Fixes the Mass-Radius Relation Characterizing Different Polytropes} %p11

We consider only Emden functions, which are regular at the origin and normalized to $\theta_n(0)=1$,
$\theta_n^{\prime}(0)=0$. Their first zeros $\theta(\xi_{1n})=0$ determine the stellar radius $R=\alpha\xi_{1n}$.
Each Emden function of index $n$ is characterized equivalently by its dimensionless outer radius $\xi_{1n}$, its outer
boundary value $_0\omega_n$ of the homology invariant $\omega_n$, or its density ratio $\rho_{cn}/\bar\rho_n$, where $\bar\rho_n=3M/4\pi R^3$ is the mean density. All three are
tabulated in the third to fifth columns of Table~II, for eight values of the polytropic index $n$. Scaling relates the
mass and radius, according to the $M$-$R$ relation $M\sim~_0\omega_n R^{(n-3)/(n-1)}$ in the last column.

We define the {\em inner core radius} $\xi_{{\rm ic}n}$ implicitly by $u(\xi_{{\rm ic}n})=2$, the radius where
the acceleration
$Gm/r^2$ reaches a maximum and the gravitational energy density overtakes the internal energy density. The sixth and
seventh columns in Table~II list dimensionless values for this core radius $r_{{\rm ic}n}/R=\xi_{{\rm ic}n}$ and
included mass $m_{{\rm ic}n}/M$, shown by red dots in Figures~3,~4.
According to equation~(\ref{eq:energyratio}), the internal energy dominates in the core; while in the envelope,
the gravitational energy dominates.

\begin{table*}[b] %%%%Table II p11
\caption{Scaling Exponents, Core Parameters, Surface Parameters, and Mass-Radius Relations for Polytropic Gas Spheres
of Increasing Core Concentration}
\begin{ruledtabular}
\begin{tabular}{|l|l||l|l|l||l|r||r|}
$n$ &$\tilde{\omega}_n$ &$\xi_{1n}$ &$\rho_{cn}/\bar\rho_n$&$_0\omega_n$ &$r_{{\rm ic}n}/R$&$m_{{\rm ic}n}/M$
&$M\sim R^{1-\tilde{\omega}_n}$ Properties  \\
\hline %\hline
0   &-2         &2.449              &1                  &0.333          &1         &1            &$M\sim R^3$; incompressible matter, all core\\
1   &$\pm\infty$&3.142              &3.290              &...            &0.66      &0.60         &$R$ independent of $M$ \\
1.5 &4          &3.654              &5.991              &132.4          &0.55      &0.51         &$M\sim R^{-3}$; nonrelativistic degenerate\\
2   &2          &4.353              &11.403             &10.50          &0.41      &0.41         &                     \\
3   &1          &6.897              &54.183             &2.018          &0.24      &0.31         &$M$ independent of $R$; Eddington standard model\\
4   &2/3        &14.972             &622.408            &0.729          &0.13      &0.24         &                      \\
4.5 &4/7        &31.836             &6189.47            &0.394          &0.08      &0.22         &                 \\
5   &1/2        &$\infty$           &$\infty$           &0              &0         &0.19          &maximally concentrated; entirely envelope \\
\end{tabular}
\end{ruledtabular}
\end{table*}

For homology variables, we prefer a new independent variable $z:=3-u=-d\log{\bar{\rho}_n}/d\log{r}$ and a new dependent
variable
$w_n:= nv_n:=-d\log{\rho}/d\log{r}$. In term of these variables, the hydrostatic equilibrium and of mass continuity
characteristic equations~(\ref{eq:chareqns}) are
\be d z/(3-z)(w_n-z)=d\log{w_n}/(2-z+w_n /n)=d \log{r}\quad .\ee
The first equality is the first-order Abel equation for the invariant $w_n(z)$, which we solve for the central boundary
condition $w_n \rightarrow 5 z/3$ when $z\rightarrow 0$. $w_n(z)$ and the differences $(3/5)[w_n(z)-w_5(z)]=
(3/5)w_n (z)-z=d \log{ \bar{\rho}/\rho^{\frac{3}{5} }} \sim G_n$ are plotted in Figures~2 and~1, respectively,
for polytropic indices $n=1,2,3,4,5$.

\begin{figure}[t] %%%%Fig. 3 p12
\includegraphics[scale=0.65]{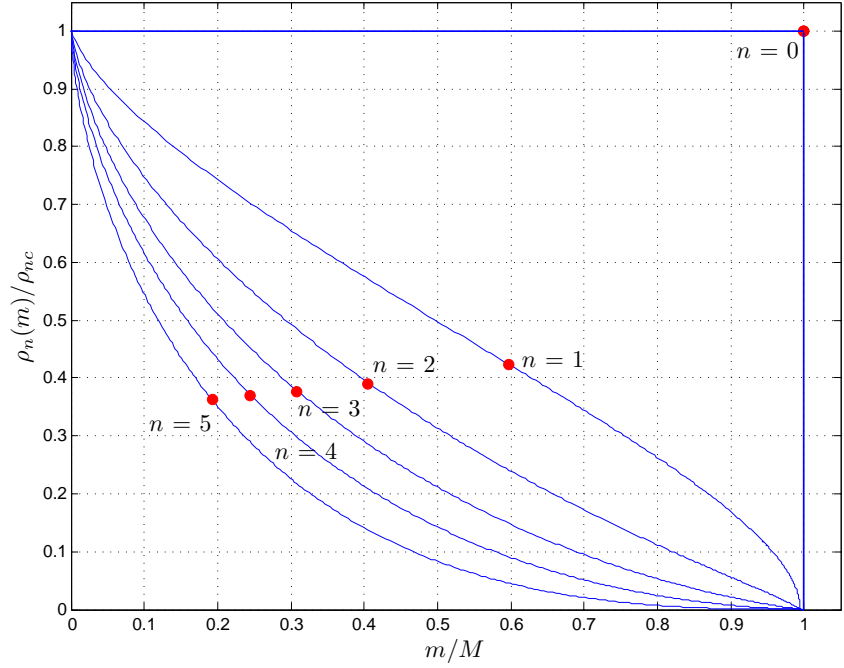}
\caption{Normalized density profiles as a function of fractional included mass $m/M$, for polytropes of finite mass $M$
and compressibility increasing with $n$. The red dots mark the cores. For incompressible matter ($n=0$), the polytrope
is all core. As the matter softens ($n$ increases), an envelope grows to ultimately encompass just over 80\% of the
mass. For any $n>1$, the density at the inner core radius stays in the narrow range
$0.37 < \rho(r_{{\rm ic}n})/\rho_c < 0.42$.}
\end{figure}

\begin{figure}[t] %%%%Fig. 4 p13
\includegraphics[scale=0.65]{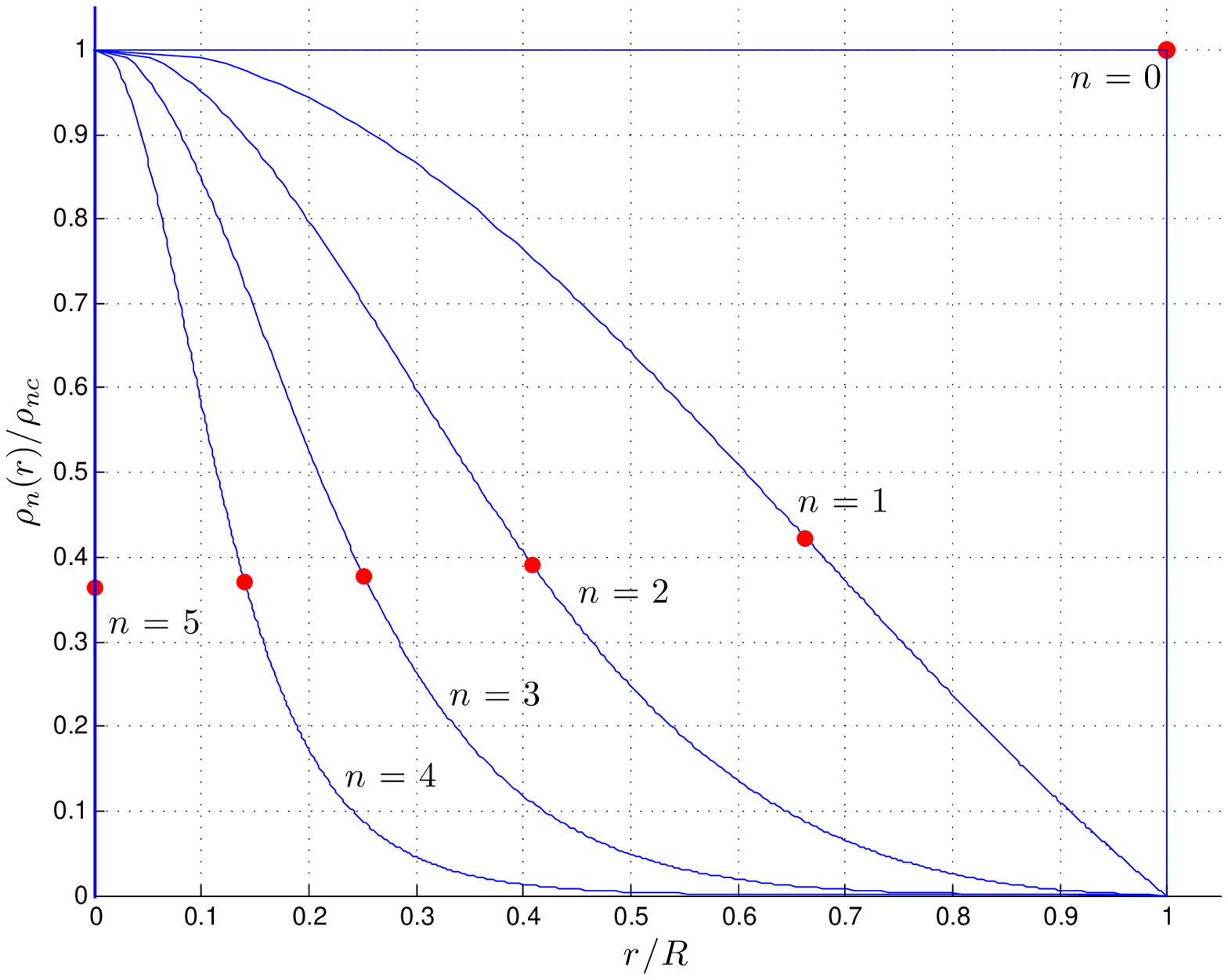}
\caption{Normalized density profile as function of fractional radius $r/R$. The density is constant for incompressible
matter ($n=0$) which is all core, but is concentrated at the origin for an unbounded polytrope ($n=5,~R=\infty$)
which is all envelope.  For any $n>1$, the density at the inner core radius stays in the narrow range
$0.37 < \rho(r_{{\rm ic}n})/\rho_c < 0.42$.}
\end{figure}

For $n<5$, the stellar boundary lies at finite radius.  The Noether charge is
nearly conserved at $G_n\approx 0$ in the inner core, but grows rapidly as the boundary radius is approached
(Figure~1).  In the envelope, the growing Noether charge measures how close the boundary is.  The finite radius $R$ determines the stellar scale, even though the polytropic form is locally scale invariant.

For incompressible matter ($n=0$), there is no core concentration: the mass is uniformly distributed, and the entire
star is core.  But as the equation of state softens as $n$ increases toward $5$, the gradient
decreases, the core concentrates, the inner core radius shrinks, and the envelope outside the core grows:
$r_{{\rm ic}n}/R\rightarrow 0$, $m_{{\rm ic}n}/M\rightarrow \sim 0.19$.
For the softest equations of state $n\lesssim 5$, the stellar radius $\xi_{1n} \approx 3(n+1)/(5-n)$, the inner core
radius shrinks $\xi_{{\rm ic}n}\approx \sqrt{10/3 n}$, their ratio $r_{{\rm ic}n}/R=\xi_{{\rm ic}n}/\xi_{1n} \approx
0.045(5-n)$, $m_{{\rm ic}n}/M \approx 0.20$, and $_0\omega_n\approx\sqrt{3/\xi_{1n}}$.

For $n=5$, the core becomes infinitely concentrated, shrinking to zero, and
the star is all envelope.  The $n=5$ regular solution
\be \theta_5(\xi)=(1+\xi^2 /3)^{-1/2}\quad ,\quad \rho=\rho_c (1+\xi^2 /3)^{-5/2} \quad ,
\quad m=M\xi^3/(3+\xi^2)^{3/2}\quad , \quad v_5=(3-u)/3 \ee
has infinite stellar radius $R$ as shown at the bottom of Table~II.
In this case, not only is the differential form scale invariant,  but also the action and stellar structure.
The $n=5$ polytrope is globally scale invariant, and the Noether charge $G_5 \sim (v_5+u/3-1)=0$.

For $0<n<5$,
\be w_n (z)=\int_0 ^z\ dz\ w_n\frac{(2-z+w_n /n)}{(3-z)(w_n-z)}\approx (5/J)[1-(1-z/3)^J]:=w_{n{\rm Pic}}(z)\quad ,
\quad J:=(9n-10)/(7-n)\quad ,\ee
is well-approximated by the {\em Picard approximation} obtained by inserting the core values $w_n (z)\approx$
$(5/3)z$ inside the integrals. Indeed, this Picard approximation is everywhere exact
for $n=0,~5$. For $0<n<5$,
it breaks down only in the outer envelope, where $w_n$ diverges as $w_n \rightarrow n[_0\omega_n ^{n-1}/u]^{1/n}$, and
$_0\omega_n$ must be calculated from the exact asymptotic value of $\omega_n$ given in Table~II.

Integrating over $z$, the density profile and Emden functions are ~\cite{BludKen}
\bea
\rho_n(z)/\rho_{cn}=\theta_n ^n =\exp{\Bigl\lbrace -\int_0 ^z  \frac{dz\ w_n(z)}{[w_n(z)-z](3-z)} \Bigr\rbrace} \approx
(1-z/3)^{5/2} \\
\theta_n = \exp{\Bigl\lbrace  -\int_0 ^z  \frac{dz\ w_n(z)}{n [w_n(z)-z](3-z)}\Bigr\rbrace} \approx
(1-z/3)^{5/2n}:=\theta_{n{\rm Pic}}\quad  ,\eea
where again the Picard approximations are obtained by inserting the core relations
$w_n(z)=n v_n(z) \approx (5/3)z$ under the integrals.

\begin{table*}[t] %%%%Table III p12
\caption{Taylor Series and Picard Approximations $\theta_{n{\rm Pic}}$ to Emden Functions $\theta_{n}$}
\begin{ruledtabular}
\begin{tabular}{|l||l||l|l}
$n$  &Emden Function and Taylor Series &$N:=5/(3n-5)$ &Picard Approximation $\theta_{n{\rm Pic}}:=
(1+\xi^2/6N)^{-N}$  \\
\hline \hline
0    &$1-\xi^2/6$                                   &-1         &$1-\xi^2/6$                              \\
1    &$\sin{\xi}/\xi=1-\xi^2/6+\xi^4/120-\xi^6/5040+\cdots$&-5/2&$(1-\xi^2 /15)^{5/2}=1-\xi^2/6+\xi^4/120-\xi^6/10800+\cdots$    \\
$n$  &$1-\xi^2/6+n \xi^4/120-n(8n-5)/15120 \xi^6+\cdots$   &$5/(3n-5)$ &$(1+\xi^2/6N)^{-N}=1-\xi^2/6+n \xi^4/120-n(6n-5) \xi^6 /10800+\cdots$ \\
5    &$(1+\xi^2/3)^{-1/2}$                                 &1/2        &$(1+\xi^2/3)^{-1/2}$ \\
\end{tabular}
\end{ruledtabular}
\end{table*}

\section{CLOSED FORM APPROXIMATION TO EMDEN FUNCTIONS}  %%V p11

With the solutions $w_n(z)$ to the first-order equation, we now use the second equation~(\ref{eq:chareqns})
\be dm/m:=u\cdot dr/r=dz/[w_n(z)-z] \ee
to obtain
\bea
m(z)/M=(z/3)^{3/2}\cdot \exp{\Bigl\lbrace \int_0 ^z dz \Bigl\lbrace
\frac{1}{[w_n(z)-z]}-\frac{3}{2z}\Bigr\rbrace \Bigr\rbrace} \approx (z/3)^{3/2}\\
r(z)/R=(z/3)^{1/2}\cdot \exp{\Bigl\lbrace  \int_0 ^z  dz \Bigl\lbrace \frac{1}{(3-z)[w_n(z)-z]}-
\frac{1}{2z}\Bigr\rbrace \Bigr\rbrace } \approx (3z)^{1/2}/(3-z)\quad  \eea
for the mass and radial distributions.
The integration constants $R,~M,~\rho_c$ express the scale dependence of the polytrope.

Using the radial distribution (58) to eliminate $z(\xi)$, the Picard approximations
\be \theta_{n{\rm Pic}}(\xi)=(1+\xi^2/6N)^{-N}, \qquad N:=5/(3n-5) \ee
to the Emden functions are obtained and tabulated
in the last column of Table~III. For $n=0$ and 5 polytropes, this closed form is exact.
For intermediate polytropic indices $0<n<5$, the Picard approximation breaks down near the outer boundary, but remains a good approximation over most of
the polytrope's bulk.

Indeed, the Picard approximation is far better than any truncation of the Taylor series expansion of $\theta_n$, whose radius of convergence
is $\xi\approx 2$.
For the worst case, the Eddington standard model ($n=3$), the Picard approximation
$\theta_{3{\rm Pic}}(\xi)$ to the exact Emden function and its tenth-order polynomial approximation:
\be
\theta_3(\xi)\approx 1-\xi^2/6+\xi^4 /40-(19/5040) \xi^6+(619/1088640) \xi^8 -(2743/39916800) \xi^{10}\quad
\ee
are shown in Figure 5. Because this Picard
departs from the Taylor series expansion already in sixth order
\be \theta_{3{\rm Pic}}(\xi)=(1+2 \xi^2/15)^{-5/4}=\theta_3+\xi^6 /3528 +\cdots \quad , \ee
it remains $90\%$ accurate out to $\xi\approx 3.5$, more than twice the core radius and more than half-way out to
the stellar boundary at $\xi_{13}=6.897$.  Except for their very outer envelopes, which contain little mass and
are never polytropic, the Picard approximations in white dwarf and ZAMS stars should be even better than for this $n=3$ polytrope.

 \begin{figure}[t] %%%%Fig. 5 p14
\includegraphics[scale=0.65]{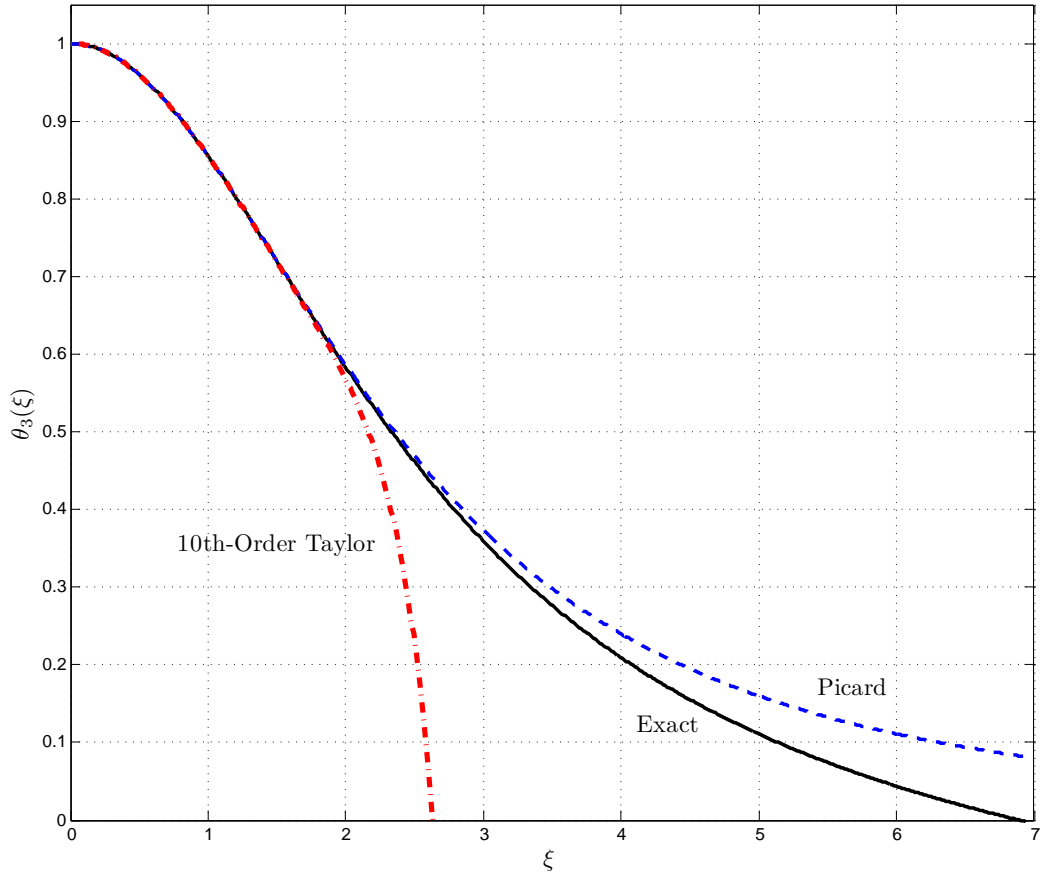}
\caption{The exact Eddington standard model Emden function $\theta_3(\xi)$, its Taylor series truncated at tenth-order, and its Picard
approximation. Even in this worst case, the Picard approximation works well outside the core radius at
$\xi_{{\rm ic}3}=\xi_{13}(r_{{\rm ic}3}/R)$ = 1.65, but breaks down near the boundary. For nonstandard polytropes ($n\neq 3$), the Picard
approximation is even better and becomes exact everywhere as $n \rightarrow$ 0 or 5.}
\end{figure}

\section{CONCLUSIONS} %%VI p14

We have generalized Noether's theorem connecting variational symmetries to conservation laws to generalized symmetries
of the Euler-Lagrange equations.  Although these lead only to nonconservation laws, they still reduce the Euler-Lagrange
equations to first order, plus a quadrature. For scaling symmetries, the nonconservation law takes a special form
linearly connecting the ``kinetic'' and ``potential'' parts of the Lagrangian.  In special cases, a symmetry of the
Euler-Lagrange equations and the nonconservation law can reduce to a conservation law and symmetry of the Lagrangian,
the action, and possibly the solution.

For nonrelativistic systems with inverse power law potentials, the scaling nonconservation law is a Lagrange's identity,
leading to generalized virial theorems.  For spherical hydrostatic systems obeying barotropic equations of state,
the scaling nonconservation law leads to an analogous linear relation between the local gravitational and internal
energies.  From this nonconservation law, we derive all the properties of polytropes. Quadrature then leads to the
regular (Emden) functions and their Picard approximations, which are useful wherever stars are approximately or exactly
polytropic.

\appendix
\section{LAGRANGIAN FORMULATION OF BAROTROPIC HYDROSTATICS}  %%%p14

Stellar structure generally depends on coupled equations for pressure equilibrium and heat transport.  Only if the heat
transport leads to a local {\em barotropic} relation $P=P(\rho)$ can the hydrostatic equations be considered
independently. In such {\em barotropes}, the mechanical structure is fixed without reference to the thermal structure

\subsection{Mass Continuity and Hydrostatic Equilibrium}

We consider a self-gravitating isolated system in local thermodynamic equilibrium, a barotrope held at zero external
pressure. The thermodynamic potential energy
or work need to adiabatically extract unit mass is the {\em specific enthalpy} $H(\rho)=E+P/\rho$.  Barotropic energy
conservation, $d H:=d P/\rho$, makes the specific enthalpy a more natural state variable than the specific internal
energy $E$, pressure $P$, or density $\rho$. The equation of hydrostatic equilibrium $-d P/dr=Gm\rho /r^2:=\rho g$ is
then
\be -d H/dr=dV/dr = g = Gm(r)/r^2 \quad , \label{eq:hydrostaticeq-1}\ee
describing how this local specific enthalpy or extraction energy $H(r)$ depends on the local gravitational potential
$V(r)$. Integrating, we have the energy conservation equation
\be H(r)+V(r)=-GM/R\quad ,\quad r<R \quad ,\ee
where the zeros of the gravitational potential and specific enthalpy have been chosen at infinity and at the spherical
surface, respectively.

Because the gravitational potential obeys Poisson's equation
\be \nabla^2 V=\frac{1}{r^2}\frac{d}{dr}\Bigl( r^2\frac{dV}{dr}\Bigr)=4\pi G \rho \quad ,\ee
the specific enthalpy obeys the second-order equation
\be \fbox{$ \displaystyle \nabla ^2 H+4\pi G\rho (H) =0 \label{eq:hydrostaticEL-2}$} \quad .\ee

Implementing the equation of hydrostatic equilibrium requires a local entropic relation $P(\rho)$, or $P(H)$,
$\rho (H)$, which is determined by the thermal stratification
of the static matter distribution in local thermodynamic equilibrium, and by a central boundary, or regularity,
condition $(d P/d r)_0=0=(d \rho/d r)_0$. Near the origin,
\bea
\rho\rightarrow\rho_c (1- B r^2)\quad , \quad m(r)\rightarrow\frac{4 \pi \rho_c r^3}{3}\Bigl( 1-\frac{3}{5}B r^2\Bigr)
\quad ,\nonumber \\
\bar{\rho}(r):= 3 m(r)/4 \pi r^3 \rightarrow\rho_c \Bigl( 1-\frac{3}{5} B r^2\Bigr) = \rho_c^{2/5} \rho^{3/5}\quad ,
\quad dw/du \rightarrow -5/3 \quad ,
\eea
in terms of the homology variables $w:=-d \log{\rho}/d\log{r},~u:=d\log{m}/d\log{r}$
for the mass density and included mass.

\subsection{A Constrained Minimum Energy Principle for Hydrostatic Equilibrium}

In a static, self-gravitating sphere of mass $M$ and radius $R$, the {\em Gibbs free energy}
\be W:=E-TS+PV=\Omega+U\quad , \ee
in terms of the gravitational and internal energies
\be \Omega=-\int_0^M (Gm/r) dm\quad , \quad U=-\int_0^R P d\mathcal{V}\quad , \ee
where $\rho$, $E$, and $-Gm(r)/r$ are the mass density, specific internal energy, and gravitational potential,
respectively.
In the Eulerian description, the radial coordinate is $r$, the enclosed volume is $\mathcal{V}=4\pi r^3/3$, and
the enclosed mass $m(r)$ is constrained by mass continuity $d m(r)=\rho d \mathcal{V}$.
The Gibbs free energy is the work available in adiabatically expanding the sphere at fixed external pressure.  If
\be W = \int_0^R \mathcal{L}(r,m,m') dr = -\int_0^R 4\pi r^2[Gm\rho /r+P(\rho)] dr \quad ,\ee
abbreviating $':=d /dr$, then the Lagrangian $\mathcal{L}$ is the Gibbs free energy per radial shell $dr$.

The constrained minimum energy variational principle~\cite{Hansen,Chui} for hydrostatic equilibrium is that
the Gibbs free energy
be stationary ($\delta W=0$) under adiabatic deformations in specific volume $\delta \mathcal{V}_{\rho}=
d(4\pi r^2\delta r)/dm$ that vanish on the boundaries and satisfy the mass continuity constraint $m' = 4\pi r^2\rho$.
This minimum energy principle has the equation of hydrostatic equilibrium
\be \mathcal{D}_r:=Gm/r^2 + H'=Gm/r^2+dP/\rho dr = 0\quad ,\ee
as its Euler-Lagrange equation, with mass continuity as a constraint.  This equation is scale invariant if the specific
enthalpy $H$ scales as $m'$.

\subsection{An Unconstrained Variational Principle} %p15

Using Poisson's equation to incorporate the mass continuity constraint, the gravitational energy is
\be \Omega=-\int_0 ^R (V'^2/2) 4 \pi r^2 dr \quad ,\ee
so that the second-order Lagrangian (used in Section~IV)
\be \mathcal{L}(r,H,H')=4\pi r^2[-H'^2/8\pi G+P(H) ]\ee
is unconstrained and has Euler-Lagrange equation~(\ref{eq:hydrostaticEL-2}).  The canonical momentum and Hamiltonian are
\be p:=\partial\mathcal{L}/\partial H'=-r^2 H'/G = -m\quad ,\quad
\mathcal{H}(r,H,p)=-Gp^2/2 r^2-4\pi r^2 P(H)\quad , \label{eq:hydrostaticHamilton} \ee
and the canonical equations are
\be \partial\mathcal{H}/\partial p=H'=-Gp/r^2\quad ,
\quad \partial\mathcal{H}/\partial H=-p '=m'=4\pi r^2\rho \quad .\ee
Spherical geometry makes the system nonautonomous, so that $dH/d r=-\partial\mathcal{L}/\partial r =-2\mathcal{L}/r$
vanishes only at large $r$, with vanishing sphericity.

\section{STELLAR THERMODYNAMICS AND CONVECTIVE STABILITY}

The structure of luminous stars depends upon the coupling between hydrostatic and thermal structure through an equation
of state $P=P(\rho, T,\mu)$, which generally depends on the local temperature and chemical composition.  But, ignoring
evolution, the matter entropy is locally conserved, so that steady-state stars are in both local thermal equilibrium and
mechanical equilibrium. In a fluid held in pressure equilibrium at constant external
temperature, the specific Gibbs free energy $H-TS=-V(r)$ is a minimum.
Under hydrostatic equilibrium, the density $\rho(r)$, specific
internal energy $E(r)$, specific enthalpy $H(r)=E+P/\rho$, specific entropy and thermal gradient
$\nabla (r):=d \log{T}/d \log{P}$ depend implicitly on the gravitational potential $V(r)$.

In the first law of thermodynamics
\be T dS= d Q=C_V dE+P d(1/\rho)\quad ,\ee
$E,~\rho$ can be written as functions of temperature and pressure.  Clever use of thermodynamic identities then
leads to~\cite{Hansen,Kippen}
\be TdS=C^{*} dT\quad ,\quad d S=C_P(\nabla-\nabla_{ad}) d\log{P}\quad ,\ee
where the {\em gravithermal specific heat} $C^{*}:=d S/d\log{T}=C_P(1-\nabla_{ad}/\nabla )$
depends on the specific heat $C_P$ and on the {\em adiabatic gradient} $\nabla_{ad}:=
(\partial \log{T}/\partial \log{P})_S$.
This expression of the first law of thermodynamics
relates the local {\em thermal gradient} $\nabla (r):=d \log{T}/d \log{P}$
to the gradient of the specific entropy $S(r)$,
which derives ultimately from the heat transport and generally varies in stars that are not in convective equilibrium.

According to Schwarzschild's minimal entropy production criterion, convective stability requires $dS/d\log{P}\leq 0$, so
that the specific entropy is constant in convective equilibrium and increases outward in
radiative equilibrium. This makes barotropic stars of mass $M$ extremal in two respects:
the central pressure is minimal in barotropic stars of a given radius $R$; the central pressure and temperature
are maximal in barotropic stars of given central density. Because stellar evolution is driven by developments in
the core, these bounds drive stars toward uniform entropy in late stages of evolution~\cite{Kovetz}.

\begin{acknowledgments}
SAB thanks Romualdo Tabensky (Universidad de Chile) for helpful discussions of Lagrangian dynamics and
acknowledges support from the Millennium Center for Supernova Science through grant P06-045-F funded
by Programa Bicentenario de Ciencia y Tecnolog\'ia de CONICYT and Programa Iniciativa Cient\'ifica Milenio de
MIDEPLAN. The figures were generated with MATLAB~7.
\end{acknowledgments}

\bibliography{bibliographyLE} %Data base makes ScaleInvarianceSynthesis.bbl needed for references.
\end{document}